\documentclass[12pt]{article}
\usepackage{amssymb} 
\def\IP{\relax{\rm I\kern-.18em P}}

\newcommand{\nc}{\newcommand}
\newcommand{\beq}{\begin{equation}}
\newcommand{\eeq}{\end{equation}}
\newcommand{\beqa}{\begin{eqnarray}}
\newcommand{\eeqa}{\end{eqnarray}}
\def\o{\otimes}

\def\comments#1{}

\def\QC{\mathbb{C}}

\def\QZ{\mathbb{Z}}

\def\tr{{\rm tr\ }}

\def\ket#1{|#1\rangle}

\def\kket#1{|#1\rangle\rangle}

\def\CN{{\cal N}}

\def\mathbfn#1{{\mbox{\boldmath $ #1 $}}} 
\def\smathbfn#1{{\mbox{\boldmath $\scriptstyle #1 $}}}

\def\a{\alpha}

\def\II{\relax{I\kern-.10em I}}

\def\IZ{\relax\ifmmode\mathchoice
{\hbox{\cmss Z\kern-.4em Z}}{\hbox{\cmss Z\kern-.4em Z}}
{\lower.9pt\hbox{\cmsss Z\kern-.4em Z}}
{\lower1.2pt\hbox{\cmsss Z\kern-.4em Z}}\else{\cmss Z\kern-.4em
Z}\fi}

\font\cmss=cmss10 \font\cmsss=cmss10 at 7pt
\def\IR{\relax{\rm I\kern-.18em R}}

\def\BZ{\QZ} 
\def\BP{\IP}
\def\BC{\QC}

\def\lp10{l_P^{10}}
\def\lp11{l_P^{11}}
\def\R11{R_{11}}

\def\frac#1#2{{#1 \over #2}}

\nc{\nn}{\nonumber}
\def\lb{\left( }
\def\href#1#2{#2}
\def\rb{\right) }
\def\IR{\relax{\rm I\kern-.18em R}}
\def\IZ{\relax\ifmmode\hbox{Z\kern-.4em Z}\else{Z\kern-.4em Z}\fi}
\def\IP{\relax{\rm I\kern-.18em P}}

\def\mathR{{\bf R}}
\def\mathZ{{\bf Z}}
\def\proj{{\mbox{\rm \scriptsize proj}}}
\def\W{{\cal W}}
\def\J{{\cal J}}
\def\H{{\cal H}}
\def\cH{{\cal H}}
\def\S{{\cal S}}
\title{\bf D-branes at Singular Curves of \\[2mm] 
  Calabi-Yau Compactifications\\[5mm] }
\author{{\sc Ilka Brunner} \\[2mm] 
 Dept.\ of Physics and Astronomy, Rutgers University, \\
 Piscataway, NJ 08855, U.S.A.  \\[5mm]  
     {\sc   Volker Schomerus} \\[2mm]
II. Inst.\ f\"ur Theoretische Physik, Universit\"at Hamburg
\\ Luruper Chaussee 149, D--22761 Hamburg, Germany}
\date{January 20, 2000}
\begin{document}
\begin{titlepage}      \maketitle       \thispagestyle{empty}

\vskip1cm
\begin{abstract}
\noindent  
We study the Gepner model description of D-branes in
Calabi-Yau manifolds with singular curves. From a geometrical
point of view, the resolution of singularities leads to
additional homology cycles around which branes can wrap.   
Using techniques from conformal field theory we address 
the construction of boundary states for branes wrapping 
additional 3-cycles on the resolved Calabi-Yau manifold. 
Explicit formulas are provided for $\BZ_2$ singular 
curves. 

\end{abstract}
\vspace*{-18.9cm}
{\tt {DESY 00-010 \hfill RUNHETC 2000-02 }}\\
{\tt \phantom{{UUITP-nn/99} \hfill hep-th/0001132}}
\bigskip\vfill
\noindent
\phantom{wwwx}{\small e-mail:}{\small\tt ibrunner@physics.Rutgers.EDU; 
vschomer@x4u.desy.de} 
\end{titlepage}
\section{Introduction}
During the last years our understanding of non-perturbative
aspects of string- and field theories has improved tremendously.
In this process, the analysis of D-branes played a prominent role. 
While many of these investigations have been 
devoted to string theories in a flat background, 
it has been tried to extend this knowledge to curved backgrounds
preserving fewer supersymmetries using various approaches. 
In this context, Calabi-Yau (CY) compactifications are of
particular interest. There
exist essentially two ways to approach $D$-branes in CY-manifolds: 
in the large radius limit we can describe them geometrically as 
branes wrapping around holonomy cycles \cite{BeBeStr,OoOzYi}. 
It has been found that D-branes in a geometrical phase
are naturally described in terms of K-theory classes \cite{MiMo, WittK}.
On the  other hand, these geometrically intuitive concepts
are not available in the stringy regime. Here,
we can employ methods of boundary conformal field 
theory (CFT) instead to study D-branes at Gepner points \cite{RSgepner,
GutSat1,GutSat2,NakNoz}. The comparison of these two approaches 
has been initiated in \cite{BDLR} (see also \cite{DiaRom,KLLW,
Scheid} for more recent work in this direction). 
\smallskip
 
The present work focuses mainly on extending the CFT constructions
of D-branes at the Gepner points of CY compactifications. While a 
large class of such D-branes was obtained in \cite{RSgepner}, a 
closer investigation of their RR-charges shows that none of them
corresponds to branes wrapping the exceptional cycles which appear 
from the resolution of possible singularities. Since there are only 
a few non-singular cases (of which the quintic is the most 
prominent), one would certainly hope that CFT-techniques can 
provide additional D-branes for almost all Gepner models. This
is the problem we are going to address below. In particular, we 
will construct so-called boundary states that are associated 
with branes wrapping the additional three-cycles induced by
the resolution of  a $\BZ_2$-singularity
over a curve in the CY-space. 
\medskip

D-branes in a flat space with orbifold singularities have been
studied extensively in the string theory literature, starting
with the work of Douglas and Moore \cite{DM}. In particular,
it is known how to describe D-branes wrapping the collapsed
cycles at the orbifold point.
The main idea 
is to attach Chan-Paton factors to the ends of the open strings and define
an orbifold action on them. Boundary states corresponding to
branes away from the orbifold fixed points are obtained by
summing over the brane's pre-images in the covering space.
At the fixed points, however, the expressions for boundary 
states can involve contributions from twisted sectors of the
theory, leading to a charge under RR-potentials coming from
the twisted sector. We shall see these concepts reappearing in the non-%
geometrical CFT analysis.
\smallskip 

Our analysis of D-branes in Gepner models will be formulated in 
the framework of simple current orbifolds. For closed string 
theories, the necessary techniques were developed long ago in 
\cite{syone,sytwo}. Open string descendants of simple current 
orbifolds were investigated systematically by Fuchs and 
Schweigert  \cite{FSshort,FSI,FSII,FSwzw} and more recently in
\cite{schellopen}. Some non-trivial examples have also been 
studied previously in \cite{PrSaSt1,PrSaSt2}. The general 
results show that the boundary states of \cite{RSgepner} 
can be further `resolved', if the action of an appropriate set 
$\Gamma$ of simple currents on conformal families of the theory 
possesses short orbits, i.e.\ orbits of length being less than 
the order $|\Gamma|$ of $\Gamma$. 
\smallskip

Some general aspects of simple current orbifolds will be
discussed in the next section. In particular, we shall 
explain how one can obtain a large set of D-brane states 
in the orbifold theory by an appropriate projection. For reasons
to become clear later, the branes that result from this 
construction will be called {\em untwisted} D-branes. 
In Section 3 we illustrate the whole procedure at the 
example of A-type boundary states in Gepner models, thereby 
recovering precisely the A-type D-brane states listed in  
\cite{RSgepner}. Geometrically, these correspond to branes 
wrapping middle dimensional cycles on the CY-manifold 
\cite{OoOzYi}. As we argue in Section 4, the untwisted 
D-branes do not wrap exceptional cycles. This motivates 
our search for additional D-brane states in Section 5. 
There we will show that untwisted D-branes at a $\BZ_2$ 
singularity over some curve $C$ can be further resolved. 
Explicit formulas for the associated boundary states and
the open string partition functions are provided. We 
finally conclude with a number of remarks on possible 
extensions. These include the analysis of B-type boundary
states and of branes wrapping $\BZ_N$-singularities. While
the present techniques do not suffice to resolve branes at 
$\BZ_{N\neq 2}$-singularities, they can be used to study 
B-type boundary states and the comparison with the results
of \cite{RSgepner} provides new evidence for mirror 
symmetry in the open string sector.

\section{Orbifolds and untwisted D-branes}
\def\orb{{\mbox{\rm \scriptsize orb}}}

The aim of this section is to review some results on simple current 
orbifolds and a general method for the construction of D-branes in 
the untwisted sector of the orbifold theory. This will be applied to
Gepner models below.

\subsection{Simple current orbifolds - the bulk theory}  
The simple current techniques developed in \cite{syone,sytwo}
allow to construct new modular invariants from existing ones. A 
well-known class of examples is provided by Gepner models 
\cite{Gepn1,Gepn2}, where the GSO-projected partition function is
obtained using the spectral flow operator as a simple current.
Other applications in string theory include the construction of
$(0,2)$ models which lead to $\CN=1$ space-time supersymmetric
theories in four uncompactified dimensions \cite{BluWi}.
\medskip

In this subsection, we briefly summarize the main results of 
\cite{syone,sytwo}. Consider some given bulk theory with 
a bosonic chiral algebra $\W$. We label classes of irreducible 
representations of $\W$ by labels $i,j,k$ taken from an index set 
$\J$. Within
$\J$ we may find some non-trivial classes $g \in \J$ such that
the fusion product of $g$ with any other $j \in \J$ gives again
a single class $g \cdot j = gj \in \J$. Such classes $g$ 
are called {\em simple currents} and the set ${\cal C}$ of all 
these simple currents forms an abelian subgroup ${\cal C} \subset 
\J$. The product in ${\cal C}$ is inherited from the fusion product
of representations. From now on, let us fix some subgroup 
$\Gamma \subset {\cal C}$.
\smallskip  

Through the fusion of representations, the index set $\J$
comes equipped with an action $\Gamma \times \J \rightarrow
\J$ of the group $\Gamma$ on labels $j \in \J$. Under this
action, $\J$ may be decomposed into orbits. The space of these
orbits will be denoted by $\J / \Gamma$ and we use the symbol
$[j]_\Gamma$ to denote the orbit represented by $j \in \J$.
By construction, the length of the orbit of the identity $1
\in \J$ is given by the order $|\Gamma|$ of the group
$\Gamma$. Other orbits may be shorter since there can
be {\em fixed points}, i.e.\ labels $j \in \J$ for which
\beq\label{fp}
 g \cdot j  \ = \  j   \ \ \ \mbox{ for some } \ \ \ g \in
\Gamma \ \ .
\eeq
The subgroup of all simple currents leaving some $j \in \J$
fixed is called the {\em stabilizer of $j$}:
\beq\label{stab}
\S_{j} \ = \ \{ \ g \in \Gamma \ \mid\  g \cdot j \ = \ j\ \}
\eeq
Two labels $j_1,j_2$ possess isomorphic stabilizers, if
they are on the same orbit, i.e.\ if there exists an element
$g \in \Gamma$ such that $g \cdot j_1 = j_2$.  
\medskip

For all simple currents $g \in \Gamma$ and all labels
$j \in \J$ we define the {\em monodromy charge} $Q_g(j)
\in \mathR/\mathZ \cong S^1$ of $j$ with respect to $g$
by
\beq \label{Qdef} 
Q_g(j) \ := \ h_g + h_j - h_{g \cdot j} \ \quad \
\mbox{\rm mod} \  1\ \ .
\eeq
Here, $h_l$ denotes the non-integer part of the conformal
dimension of the conformal primary that is associated with
$l \in \J$. The meaning of $Q_g(j)$ can be easily understood
once we choose two fields $J(z)$ and $\psi(w)$ from the
conformal families $g \in \Gamma \subset \J$ and $j \in
\J$, respectively. Their operator product will give fields
$\psi'(w)$ within a single conformal family $g \cdot j$,
i.e.\   
\beq \label{OPEQ} 
J(z) \, \psi (w) \ \sim \ (z-w)^{h_J + h_\psi - h_{\psi'}}\
\psi'(w)\ + \dots \ .  
\eeq
If we move $z$ once around $w$ we pick up some phase factor
$\exp(2\pi i Q_g(j))$ which is given by the monodromy charge
defined above. Let us finally note that the map $\exp(2 \pi i
Q_{.}(j)): \Gamma \rightarrow S^1$ defined by $g \mapsto \exp
(2 \pi i Q_g(j))$ gives rise to a 1-dimensional representation
of the group $\Gamma$. 
\medskip

In the case that the simple currents have integer conformal
weight, we can include them into the chiral algebra of the
theory to form an extended chiral algebra $\W(\Gamma)$. Note
that in this case the monodromy charge $Q_g(j)$ depends only on
the equivalence class $[j]= [j]_\Gamma$ of $j \in J$ in the
space $\J/ \Gamma$ of orbits. An orbit $[j]$ is
said to be invariant, if $Q_g([j]) = Q_g(j) = 0$ for all $g
\in \Gamma$. We can now write down the partition function for
the new orbifold theory. Under certain conditions that are 
satisfied for all cases to be discussed, conformal families
of the extended algebra $\W(\Gamma)$ are labeled by  pairs
$([j]_\Gamma,\tau)$ of invariant orbits along with irreducible
representations $\tau$ of the associated stabilizer subgroup,
i.e.\ by $[j]_\Gamma$ with $Q_\Gamma ([j]) = 0$ and $\tau: \S_j
\rightarrow U(1)$. The new orbifold theory possesses a diagonal
modular invariant partition function with respect to the
extended algebra $W(\Gamma)$, i.e.\    
\beq\label{intmod}
Z^{\orb} \ = \ \sum_{[j],Q_\Gamma([j])=0} \ |\, \S_{[j]}\, | \
            |\!\sum_{g \in \Gamma/\S_j} \chi_{gj}\, |^2
\eeq
Characters appearing with a multiplicity given by the order
of the stabilizer are associated with inequivalent representations
of the larger algebra $\W(\Gamma)$. Upon restriction to the
original chiral algebra $\W$, these representations become equivalent.
The new partition function is obviously non-diagonal with respect
to the smaller algebra $\W$.

\subsection{D-branes in the untwisted sector}

In this section we shall explain how to construct D-branes in the
untwisted sector of simple current orbifolds. Starting from 
D-branes in the original theory we can obtain certain boundary states 
for the orbifold theory by a projection onto invariant sectors. 
This projection method was invented in \cite{RSgepner} to get 
GSO invariant boundary states in Gepner models. The GSO projection 
can be understood through simple current extension or orbifolds. 
The methods of \cite{RSgepner} (for A-type boundary states) can 
easily be extended to open descendants of other simple current 
modular invariants. The idea is to construct boundary states of 
the theory with a non-diagonal orbifold partition function from 
boundary states of the  original diagonal theory.
\smallskip 

So let us suppose we are given our original theory with a partition
function 
$$ Z \ = \ \sum_{j\in \J} \ |\chi_j|^2 \ \ .  $$
We are looking for associated theories on the half plane such that 
chiral fields obey the gluing condition $W(z) = (\Omega \bar W)(\bar z)$ 
all along the boundary $z = \bar z$ and for all chiral fields $W$ of 
the theory. For one special choice of the gluing automorphism $\Omega$ 
one can apply Cardy's prescription to find a complete set of elementary 
boundary states of the original theory (see e.g.\ \cite{RSgepner} for 
more details). These are given by the formula
\beq\label{cardybs}
|I\rangle \ = \ \sum_j \frac{S_{I\, j}}{\sqrt{S_{0\, j}}} \ 
|j\rangle\rangle\ \ . 
\eeq
Here, the element $|j\rangle \rangle$ is the generalized coherent 
(`Ishibashi'-) state associated with the sector $\H_j \o \bar \H_{j}$ 
of the theory. $S$ is the modular S-matrix of the rational theory 
under consideration.
\medskip

Let us now consider the orbifold of the original theory obtained 
from the action $g \rightarrow \exp(2\pi i Q_g)$ of our abelian 
group $\Gamma$ on the state space $\cH$. Here we think of the 
monodromy charge $Q_g$ as an operator on the space $\cH = 
\oplus_j \H_j\o\bar \H_{j}$ that acts by multiplication with 
$Q_g(j)$ upon restriction to the subspaces $\H_j \o \bar \H_{j}$. 
As usual, the untwisted sector of the orbifold theory contains 
the states of the original theory that are invariant under the 
action of $\Gamma$. We can map the states from $\cH$ to 
the untwisted sector of the orbifold theory with the help of 
the projector  
$$P^0 \ = \ \frac{1}{|\Gamma|} \, \sum_{g \in \Gamma}\,  
         \exp(2 \pi i Q_g)\ \ .$$
The idea in \cite{RSgepner} was to take the (appropriately 
rescaled) image of the Cardy boundary states (\ref{cardybs}) under 
$P^0$ as boundary states of the orbifold theory, i.e.\      
\beq
|I\rangle_\proj \ = \ \kappa \, P^0 \ |I\rangle \ = \ 
 \frac{\kappa}{|\Gamma|} \ \sum_{g \in \Gamma} \ e^{2\pi i Q_g}
 \sum_j \frac{S_{I\, j}}{\sqrt{S_{0\, j}}} |j\rangle\rangle\ \ .
\label{projst} 
\eeq
By construction, these states are linear combinations
of generalized coherent states in the orbifold theory. Hence, it
only remains to check that they give rise to consistent open string
spectra. To this end one may exploit the following property of 
modular S-matrices
\beq
S_{gI\, j} \ = \  e^{2\pi i Q_{g}(j) } S_{I\, j} \ \ .
\label{Sprop} 
\eeq
Insertion into our previous expression for the projected boundary 
states $|I\rangle_{\mbox{\proj}}$  yields the interesting result
\beq\label{prorb}
|I\rangle_\proj \ = \ 
 \frac{\kappa}{|\Gamma|} \  \sum_{g \in \Gamma}\,  |g I \rangle.
\eeq
This means that in a general model, the charge projected
boundary states are orbits of boundary states in the theory
before orbifolding. The formula should be regarded as an 
algebraic analogue of the geometric prescription to sum over
pre-images of a brane in the covering space. Hence, the open 
string spectra associated with these the boundary states 
$|I\rangle_{\mbox{\proj}}$ are
\beqa \label{Zproj}  
Z^{\proj}_{IJ}(q) \ & := & \ _\proj\langle \theta I|\, \tilde 
q^{\frac12(L_0 +  \bar L_0) - \frac{c}{24}} \, |J\rangle_\proj 
\nn \\[2mm]   
& = & \frac{\kappa^2}{|\Gamma|^2}\ \sum_{g,h \in \Gamma,k} \, 
N^{gI^\vee \, hJ}_k \chi_k(q)  \ = \ \frac{\kappa^2}{|\Gamma|} 
 \sum_{g,k} N^{gI^\vee\, J}_k \chi_{k}(q)\ \ .
\eeqa
Here, $\theta$ denotes the CPT operator of the bulk theory and 
$I^\vee$ is the label conjugate to $I$ (see \cite{RSgepner} for 
details). If we choose 
\beq \kappa \ = \ \sqrt{|\Gamma|} \label{kappa} \eeq
then the 
coefficients of the characters on the right hand side are 
guaranteed to be integer. But they may still possess a common 
divisor. This happens whenever $I$ or $J$ is fixed under the 
action of some element $g \in \Gamma$. If one of the labels is 
on  such a short orbit, then the sum in (\ref{prorb}) consists 
of several equal terms. For $I = J$, the number of equal terms 
is given by the order $|\S_I|$ of the stabilizer subgroup $\S_I 
\subset \Gamma$ of $I$. We shall see below that in the orbifold 
theory there exist $|\S_I|$ boundary states associated with the
label $I$ which differ by a representation of the stabilizer
subgroup $\S_I$. The projection analysis provides one linear 
combination of these $|\S_I|$ states which contains only 
contributions from the untwisted sector.

\section{Gepner models and untwisted D-branes}
\def\ti{\times}

The aim now is to apply the general theory outlined
above to an important class of examples, namely to Gepner models
\cite{Gepn1,Gepn2} (see also \cite{Gree} for a review). These 
are exactly solvable CFTs which are used to study strings moving
on a Calabi-Yau manifold at small radius \cite{wittph}. Their 
construction employs an orbifold-like projection starting from 
a tensor products of $r$ $\CN=2$ minimal models. In our 
presentation we shall assume that there are $d=1$ complex, 
transverse, external dimensions in light cone gauge and that  
the number $r$ of minimal models equals $r=5$. 

\subsection{The building blocks of Gepner models} 
 
Our main building blocks are $\CN=2$ minimal models at level $k$.  
These are SCFTs with central charge $c={3k\over{k+2}}<3$ 
\cite{bfk,zamtwo,dpyz,snam}. One can label primaries of the 
bosonic subalgebra by 3 integers, $(l,m,s)$ taking values in 
the range  
\beq
\label{standardone}
        0\leq l\leq k\ ;\ \ 0\leq |m-s|\leq l\ ;\ \ s \in 
        \{-1,0,1\}\ ;\ \ l+m+s = 0\ {\rm mod}\  2\ .
\eeq
In many respects, $l$ and $m$ behave like the familiar labels in an 
$SU(2)_k$ WZW model. The third label $s$ determines the spin structure. 
States with $s=0$ are in the NS sector while $s=\pm 1$ correspond to 
the two chiralities in the $R$ sector. The conformal 
weights and $U(1)$ charges of these primary fields can be computed
by means of the formulas 
\beqa
\label{ntwoweights}
        h^l_{m,s} & = & {{l(l+2)-m^2}\over{4(k+2)}}+
                {{s^2}\over 8}\ \mbox{\rm  mod} \ 1 \ \ , \\[2mm]
        q^l_{m,s} & = & {m\over{k+2}}-{s\over 2}\ \mbox{\rm  mod} 
            \ 2 \ .
\eeqa
There are some distinguished labels $(l,\pm l,0) 
\in \J$ in the NS sector which are associated with the $\CN=2$ (anti-) 
chiral primaries. They play a special role in the Landau-Ginzburg 
description of minimal models where they are identified with powers 
$X^l \sim (l,l,0)$ of the Landau-Ginzburg 
field $X$. 
\smallskip

Our set $\J$ of conformal families contains triples $(l,m,s)$ from 
the standard range. It will be rather convenient below to consider 
an extended set $\tilde \J$ of labels $(l,m,s) \in \{ 0, \dots, k\} 
\ti \BZ_{2k+4} \ti \BZ_4$ with the only additional constraint $l+m+s 
=$ {\it even}. For each label $(l,m,s) \in \J$ there exist 
exactly two labels in $\tilde \J$, namely $(l,m,s)$ and $(k-l,m+k+2,
s+2)$. In other words, the extended set $\tilde \J$ carries an action 
of $\BZ_2$ that maps $(l,m,s)$ to  $(k-l,m+k+2,s+2)$ such that our 
original set $\J$ is simply the quotient $\tilde \J/\BZ_2$. Passing 
from $\tilde \J$ to $\J$ is known as {\em field identification}.  
\medskip

The simple currents of an $\CN=2$ minimal model can be determined
from the fusion rules. For the label $l$ these are given by the 
usual $SU(2)$ fusion rules while both other labels add like 
representations of the abelian groups $\BZ_{2k+4}$ and $\BZ_4$, 
respectively. This implies that e.g.\ $(0,1,1)$ and $(0,0,2)$ 
are both simple currents. They are of special interest in the context 
of Gepner models and will be used to generate our simple current 
group $\Gamma$. $(0,1,1)$ is the spectral flow by $1/2$ unit and 
$(0,0,2)$ the world-sheet supersymmetry generator. $(0,0,2)$ is a 
simple current of order $2$ and can be used to group the world-sheet 
fields into supermultiplets. The order of the simple current $(0,1,1)$ 
is model dependent. To see this, we apply the current $2k+4$ times to 
the identity. This will lead us back to the identity whenever the level 
$k$ is even. Since $(0,0,2)$ is nowhere on this orbit, $(0,0,2)$ and 
$(0,1,1)$ togther generate the simple current group $\Gamma = \Gamma_k 
= \BZ_{2k+4} \ti \BZ_2$ for even level $k$. When $k$ is odd, however, 
we reach the field $(0,0,2)$ after $2k+4$ applications of $(0,1,1)$ 
and hence we have to apply the simple current $(0,1,1)$ another 
$2k+4$ times. In this case, the orbit contains the label
$(0,0,2)$ and hence our orbifold group is $\Gamma = \Gamma_k = 
\BZ_{4k+8}$ for odd $k$. 

Let us briefly describe the orbits for the action of $\Gamma_k$ on 
the set $\J$. Again, we need to treat the cases of even and odd $k$ 
separately. If $k$ is odd, the orbifold group $\Gamma_k$ acts freely 
so that all orbits have length $4k+8$. For even level $k$, however, 
we generate short orbits of length $2k+4$ whenever we start 
from a field $(l,m,s)$ with $l = k/2$ because the label $l=k/2$ is
invariant under field identification. The stabilizer for these short 
orbits is a subgroup $\BZ_2 \subset \Gamma_k$.  
\medskip

From formula (\ref{ntwoweights}) and the definition relation 
(\ref{Qdef}) we can compute the monodromy charge: 
$$ Q^k_{(0,\mu,\sigma)}(l,m,s) \ := \ \frac{m\mu}{2k+4} - 
 \frac{s \sigma}{4}
 \ \ \mbox{\rm mod} \  1\ \quad \ \mbox{ for all } \ \ \ 
 (0,\mu,\sigma) \in \Gamma_k \ \ . $$
Obviously, this expression defines a representation of our orbifold
group $\Gamma_k$ for any choice of $(l,m,s) \in \J$. 
\medskip

$\CN=2$ characters and their modular properties are described 
e.g.\ in \cite{Gepn1,Gepn2}. The characters $\chi_{l,m,s}$ of the 
representations listed above can be indexed by arbitrary labels 
$(l,m,s)$ in the extended set $\tilde \J$. In fact, we can use 
the field identification to define $\chi_{l,m,s}$ for $(l,m,s)$ 
outside the standard range (\ref{standardone}) by $\chi^l_{m,s}
=\chi^{k-l}_{m+k+2,s+2}$. For the modular S-matrix 
one has the explicit formula
\beq
S^k_{(l,m,s),(l',m',s')} \ = \ \frac{1}{\sqrt{2} (k+2)}\  
 \sin \pi \frac{(l+1) (l'+1)}{k+2} \ e^{i\pi mm'/(k+2)} 
 \, e^{-i \pi ss'/2} \label{Smat2} 
\eeq 
It is easy to check that $S$ satisfies the relation (\ref{Sprop}) 
for all $g \in \Gamma_k$. 
\bigskip

In addition to the minimal models, Gepner's constructions involve 
fermions from the external space-time sector of the theory. Their 
bosonic subalgebra is an $SO(2)_1$ current algebra. Its 
representations are labeled by $s = 0,\pm1,2 \in \BZ_4$ and
they possess the obvious abelian fusion rules. All sectors in 
this theory are simple currents and their monodromy charge is 
given by   
$$ Q^{\sf f}_\sigma(s) \ :=\ - \frac{s \sigma}{4}  \ \ \mbox{\rm 
   mod} \  1\ \quad \ \mbox{ for }  \ \ \sigma = 0,\pm 1,2 . $$
The property (\ref{Sprop}) is obeyed by the modular S-matrix 
$S^{\sf f}_{s,s'} = (1/2) \exp(- i \pi s s'/2)$ of the $SO(2)_1$ 
current algebra.   
\bigskip 

\subsection{Gepner models in the bulk} 

Let us now consider a tensor product of $r=5$ minimal models with 
levels $k_i, i = 1, \dots, r$ such that their central charges adds 
up to $c=9$. To get a full string 
theory, one needs to add a sector containing ghosts and a level 
$k=1$ current algebra that comes with the space-time sector. These 
tensor products do not give consistent string backgrounds with 4d 
spacetime SUSY. But there exists some orbifold theory of this 
tensor product theory that satisfies all the requirements of a 
consistent string background. For its description we need 
further notations. Let us introduce the following vectors  
$$
{\mathbf \lambda} := (l_1,\ldots,l_r)\quad {\rm and}\quad 
{\mathbf \mu}:= (s_0; m_1,\ldots,m_r;s_1,\ldots,s_r)
$$ 
to label the tensor product of representations $(l_j,\,m_j,\,s_j)$
of the individual minimal models and of the representations $s_0=0,2,
\pm1$ that come with the level $k=1$ current algebra. The associated 
product of characters $\chi^{l_i}_{m_i,s_i}$ and $\chi_{s_0}$ is 
denoted by $\chi^{{\mathbf \lambda}}_{{\mathbf \mu}} (q)$.  
\smallskip

Next, we introduce the special $(2r+1)$-dimensional vectors $\beta_0$ 
with all entries equal to 1, and $\beta_j$, $j=1,\ldots,r$, having 
zeroes everywhere  except for the 1st and the $(r+1+j)$th entry which 
are equal to 2. These vectors stand for particular elements in the 
group $\BZ_4 \ti \prod_i \Gamma_{k_i}$. It is easily seen that they 
generate a subgroup $\Gamma = \BZ_K \ti \BZ_2^r$ where $K:= {\rm lcm}
(2k_j+4)$. Elements of this subgroup will be denoted by $\mathbfn 
\nu = (\nu,\nu_1, \dots, \nu_r)$. The monodromy charge of a pair 
$(\mathbf \lambda,\mathbf \mu)$ is 
\beqa  Q_{\smathbfn \nu} (\mathbf \lambda,\mathbf \mu) & = & \nu  
         \beta_0 \cdot {\mathbf \mu} + \sum_{i=1}^r \ \nu_i 
         \beta_i \cdot {\mathbf \mu}\ \  \mbox{\rm mod} \  1\  
     \label{Qtens}   \\[3mm] 
\mbox{where} \ \ \ \ \ \ 
\beta_0 \cdot {\mathbf \mu} &:=& 
- {s_0\over4} - \sum_{j=1}^r {s_j\over4} 
+ \sum_{j=1}^r {m_j\over 2k_j+4}\ , \\[2mm]
\beta_j \cdot {\mathbf \mu} &:=& -  {s_0\over2} 
 -{s_j\over2} \ .
\eeqa
The orbifold group $\Gamma$ acts on the labels $\mathbf \lambda$ 
and $\mathbf \mu$ in the obvious way. There appear orbits of 
maximal length $K 2^r$ and short orbits of length $K 2^{r-1}$. 
The latter are characterized by the property that $\mathbf 
\lambda = (l_1,\dots, l_r)$ satisfy $l_i = k_i/2$ for all 
$i$ such that $2k_i+4$ is not a factor in $K/2$. 
\medskip

As we mentioned before, the complete construction requires to include 
ghosts. Since the ghost sector will not play an important role in the 
later sections, we shall constrain ourselves to some brief remarks 
that are necessary in understanding Gepner models from an orbifold point 
of view. 

When we consider the  full theory, the field generating the $\BZ_K$-%
symmetry contains a factor from the ghost sector and the space-time 
part of the spin field $S^\alpha$. In the $\pm1/2$ picture, the 
operator can be represented as
\beq\label{sf}
U(z) \ = \ e^{\pm i\frac{\phi}{2}} \, 
e^{\pm i \frac{1}{2} \sqrt{\frac{c}{3}} X}\, 
S_{\alpha}
\eeq
Here, $\phi$ denotes the bosonized super-ghost and we introduced the bosonic
field $X$ whose derivative gives the U(1) current $J(z) = \sum_{i=0}^r
J_i(z)$ in the tensor product of the $SO(2)_1$ theory with the minimal
models. The precise relation is $J(z) = i \sqrt{\frac{c}{3}} 
\partial X$. 

The operator (\ref{sf}) is a simple current and its internal part agrees 
with the simple current $\beta_0$ in the tensor product considered above. 
Since the operator (\ref{sf}) has total weight one, it can be added to 
the chiral algebra and we can use the formula (\ref{intmod}) to determine
the partition function of the orbifold theory. The formula requires to 
determine invariant orbits, i.e.\ orbits with vanishing monodromy charge. 
Taking the OPE of the spectral flow (\ref{sf}) with a vertex operator that 
represent space-time scalars of the theory gives the monodromy charge
$$
\tilde Q_{\smathbfn \nu} (\mathbf \lambda,\mathbf \mu) \ = \ \nu  
         \left( \frac{\beta_0 \cdot {\mathbf \mu}}{2} + \frac12 
         \right) + \sum_{i=1}^r \ \nu_i 
         \beta_i \cdot {\mathbf \mu}\ \  \mbox{\rm mod} \  1\  
$$
by comparison with the general formula (\ref{OPEQ}). $\tilde Q$ effectively
replaces the monodromy charge $Q$ introduced in eq.\ (\ref{Qtens}). 
The orbits of vanishing monodromy charge are those of odd integer
$U(1)$ charge. 

To write a partition function of physical states one has to 
extract the physical degrees of freedom. This can be done 
by a projection onto light-cone variables which removes, in 
particular, the ghost sector. In practice, the light-cone degrees
of freedom may be read off directly in the canonical ghost 
pictures.

An important reason to include ghosts is that the fields of the 
theory aquire the right commutation properties. One would like 
to incorporate this feature in the physical theory. Therefore, 
in the partition function, the fields are counted with a 
ghost-charge dependent phase factor $\exp (2\pi i q_{ghost})$. 
This means that the states with half-integer ghost charge, i.e.\ 
the RR-states,  contribute with a negative sign. 

Having discussed all aspects of the ghosts, which are relevant 
to the simple current construction of the Gepner partition 
function, we will discard them from now on and consider only 
the physical (light-cone) degrees of freedom. 
We are prepared to write down the partition function for a 
Gepner model describing a superstring compactification to $4$ 
dimensions. 
It is given by  
$$
Z^{(r)}_G (\tau,\bar \tau) = {1\over2} 
{({\rm Im}\,\tau)^{-{2}}\over |\eta(q)|^{2}}
\sum_{\mathbf \lambda,\mathbf \mu; \tilde Q (\mathbf \lambda,\mathbf \mu) = 0}
\ \sum_{\nu,\nu_j} \ \
(-1)^{\nu} \ \,\chi^{{\mathbf \lambda}}_{{\mathbf \mu}} (q)\ 
\,\chi^{{\mathbf \lambda}}_{{\mathbf \mu}+\nu\beta_0+\nu_1\beta_1+\ldots 
+\nu_r\beta_r} (\bar q)
$$
The sign is the usual one occurring in (space-time) fermion one-loop 
diagrams. The $\tau$-dependent factor in front of the sum accounts 
for the free bosons associated to the $2$ physical transversal 
dimensions of flat external space-time, while the $1/2$ is simply 
due to the field identification mentioned above. Except for these 
modifications,  the formula for $Z_G$ is the same as eq.\ 
(\ref{intmod}). Elements 
$g = \nu \beta_0 + \dots \nu_r \beta_r$ of the orbifold group 
$\Gamma$ are labeled by $\nu,\nu_i$ so that the second sum is 
over the full group $\Gamma$. Short orbits appear twice in the 
summation and give rise to an extra factor of $2$ which is the 
order of the corresponding stabilizer subgroup. Since our orbifold 
group $\Gamma$ is abelian, we used additive notation for the action 
of elements $g \in \Gamma$ on the labels $\mathbf \lambda, 
\mathbf \mu$.

\subsection{Untwisted D-branes in Gepner models}

Using the general formalism outlined in Section 2.2 one can 
find a large set of boundary states \cite{RSgepner} which 
respect the $\CN=2$ world-sheet algebras of each minimal 
model factor of the Gepner model separately. To this end we start 
with Cardy boundary states of the tensor product theory. They  
are given by the expression (\ref{cardybs}) along with the 
formula (\ref{Smat2}) for the modular S-matrices of minimal 
models and the simple expression for the S-matrix of 
$SO(2)_1$ that we spelled out before. The generalized 
coherent states $|j\rangle \rangle$ are now parametrized 
by pairs $j = (\lambda,\mu)$. Cardy's boundary states belong 
to some gluing condition $W(z) = \Omega \bar W(\bar z), z = \bar z,$ 
which becomes $J_{i} (z) = - \bar J_i(\bar z)$ on the U(1)-currents
of the individual theories. This means that they are A-type boundary 
conditions in the sense of \cite{OoOzYi}.  
\smallskip

The boundary states $|I\rangle =: |\Lambda, \Xi\rangle$ we have just 
described depend on a spin vector $\Lambda = (L_1, \dots, L_r)$ and 
a charge vector $\Xi = (S_0;M_1, \dots, M_r; S_1, \dots, S_r)$.
From these states in the tensor product theory we can pass to 
boundary states of the Gepner model using the general strategy 
explained in Section 2.2. The projected boundary states in the 
orbifold theory are given by (see eqs.\ (\ref{prorb},\ref{kappa}))
$$ |\Lambda,\Xi\rangle_\proj \ = \  \frac{1}{\sqrt{K 2^r}} 
   \sum_{\nu,\nu_i} \, (-1)^\nu (-1)^{\frac{\hat s_0^2}{2}} \, \, 
   |\Lambda, \Xi 
   + \nu\beta_0+\nu_1\beta_1+\ldots +\nu_r\beta_r\rangle\ \ . 
$$ 
Here, the element $\hat s_0$ is an operator acting on closed string 
states which measures the value $s_0$. The whole factor $(-1)^{\hat 
s_0^2/2}$ is needed to guarantee that in the open string partition 
function (similar to the closed string partition function) fields 
are counted with a phase factor referring to their ghost charge. 
If we insert the formula (\ref{cardybs}) with the appropriate modular 
S-matrix on the right hand side, we obtain the expressions
established in 
\cite{RSgepner},  
\beq
\label{rsstate}
       \kket{\alpha} \ := \  \kket{\Lambda,\Xi}_\proj\ =\ 
         \sum_{\mathbf \lambda,\mathbf \mu; 
      \tilde Q (\mathbf \lambda,\mathbf \mu) = 0} (-1)^{\frac{s_0^2}{2}} \ \
        B^{\lambda,\mu}_\alpha\ \kket{\lambda,\mu}\ .
\eeq
with the coefficients:
\beq
\label{rscoeff}
        B^{\lambda,\mu}_{\alpha}\ =\ {\sqrt{K 2^r}\over 2}
         \, e^{-i\pi \frac{s_0 S_0}{2}}\ 
        \prod_{j=1}^r{1\over{\sqrt{\sqrt{2}(k_j+2)}}}
                {{\sin(l_j,L_j)_{k_j}}\over
        {\sqrt{\sin(l_j,0)_{k_j}}}}\ e^{i\pi{{m_j M_j}\over{k_j+2}}}
        \ e^{-i\pi{{s_j S_j}\over{2}}}\ .
\eeq
Here $(l,l')_k = \pi (l+1)(l'+1)/(k+2)$. For these A-type boundary 
states the Ishibashi states are built on diagonal primary states, 
i.e.\ states in the untwisted sector, in accordance with our general 
theory in Section 2. The associated partition functions  
(\ref{Zproj}) aquire the following form (see also \cite{RSgepner}):
\beqa
\label{apart}
        Z_{\tilde \alpha \alpha}^A(q)& =& {1\over 2}\ \sum_{\lambda',\mu'}
        \sum_{\nu=0}^{K-1}\sum_{\nu_i=0,1}\ (-1)^{s_0' + S_0 - \tilde S_0} 
        \ \delta^{(4)} _{s_0' - \tilde S_0 + S_0 + \nu + 2 \sum\nu_i -2 }
        \nn \\[3mm] & & \hspace*{1cm} \times \   
        \prod_{j=1}^rN^{l_j'}_{L_j,\tilde L_j}
        \ \delta^{(2k_j+4)}_{\nu + M_j-\tilde
        M_j+m_j'}\ \delta^{(4)}_{s_j'-\tilde S_j + S_j + \nu + 2\nu_j}
        \  \chi^{\lambda'}_{\mu'}(q)\ . 
\eeqa 
The factor $1/2$ in front of the right hand side accounts for the fact 
that field identification causes each character to appear twice when 
we sum over $\lambda',\mu'$ taken from the extended range.  

If the two boundary conditions $\a,\tilde \a$ appearing in eq. (\ref{apart})
are both labeled by monodromy invariant orbits, they give rise to a monodromy 
invariant open string spectrum, i.e.\ to a spectrum that contains only 
odd-integer charges. One should recall, however, that non-invariant 
orbits of $\Gamma$ are also admissible as labels for boundary conditions. 
The condition for a supersymmetric open string spectrum consisting of 
monodromy invariant states is that the U(1) charge of the two orbits 
labeling the boundary conditions $\a$ and $\tilde a$ differs by an even 
integer.

\section{Singular curves on Calabi-Yau 3-folds}
 
The boundary states (\ref{rsstate}) we have constructed so far are 
only charged under $(c,c)$-fields in the untwisted sector. Typically, 
there exist additional $(c,c)$-fields in the 
twisted sectors of Gepner models. Their appearance is related to 
singularities of the associated Calabi-Yau spaces. The aim of this 
section is to explain this relation in some more detail. Once this 
is understood, it motivates the search for additional boundary 
states that are charged under fields in the twisted sectors. This 
will be addressed in the next section. Our presentation here will be 
rather sketchy. The interested reader can find more explanations and
details e.g.\ in \cite{HoKlTh,KaMoPl,Gree,LySch,FKSS}.
\medskip 

Suppose we are given a Gepner model composed from $r=5$ minimal 
models with levels $(k_1, \dots,k_5)$ chosen such that their  
central charges add up to $c=9$. Let us define integers $\omega_i 
= K / (2 k_i + 4)$ where $K = {\rm lcm}(2k_i+4)$ as before. 
Assuming that the Gepner model has an A-type modular invariant 
partition function, as we did before, we associate with it the 
following space $M= P_{(\omega_1, \dots, \omega_5)} [\frac{K}{2}]$ 
which is defined by the equation 
$$ M \ : \ \ \ z_1^{k_1+2} + \dots z_5^{k_5+2} \ = \ 0 $$ 
evaluated in a weighted projective space where $(z_1, \dots,z_5) 
\sim (\lambda^{\omega_1} z_1, \dots, \lambda^{\omega_5}z_5)$ for 
$\lambda \in \BC^*$. It is well known that the $(c,c)$-fields 
with total left- and right-moving charges $q$, $\bar q$ satisfying 
$q=1 = \bar q$ correspond to harmonic (2,1)-forms on $M$. We denote 
the space of these forms by $H^{2,1}$.    
\smallskip 

Typically, the space $M$ possesses singularities which can 
be either singular points or singular curves. In string theory
these singularities are resolved. Desingularization of singular
curves provides a non-vanishing contribution to the Hodge number 
$h^{21} = {\rm dim} H^{2,1}$. We will make a more quantitative 
statement momentarily after a brief description of the possible 
singularities. 
\medskip 

A singular curve on $M$ exists whenever three of the numbers 
$\omega_i$ have some non-trivial factor in common. To be 
specific, we suppose that these are $\omega_3,\omega_4,\omega_5$ 
and we denote their largest common divisor by $N$. Now our 
weighted projective space carries an action of $\BZ_N$ defined
by $(z_1, \dots, z_5) \rightarrow (\eta^{\omega_1} z_1, \dots, 
\eta^{\omega_5} z_5)$ where $\eta$ is some $N^{th}$ root of unity. 
By our choice of $N$, $\eta^{\omega_i} = 1$ for $i=3,4,5$ and 
hence, there is a two complex dimensional subspace in the weighted 
projective space that remains fixed under the action of $\BZ_N$. 
This subspace is parametrized by classes of $(0,0,z_3,z_4,z_5)$. 
The surface $M$ intersects with this singularity along the   
curve given by the equation 
$$    C\ : \ \ \ z_3^{k_3+2} + z_4^{k_4+2} + 
               z_5^{k_5+2} \ = \ 0  \ \ . $$
Putting all this together we see that $M$ possesses a $\BZ_N$
singularity over each point of the curve $C$. At a generic point, 
the space transverse to the singular curve looks like the quotient 
$\BC^2/\BZ_N$ where $\BZ_N$ acts on points $(w_1,w_2) \in \BC^2$ 
by $(w_1,w_2) \rightarrow  (\eta w_1, \eta^{-1} w_2)$. 

The singularity at the origin of $\BC^2/\BZ_N$ is known as a {\em 
rational double point} of type $A_{N-1}$. It is resolved by gluing 
in a chain of $N-1$ projective spaces $\BP^1$ which intersect 
pairwise transversely in one point and have self-intersection 
numbers $-2$. In other words, the intersection matrix equals 
the negative Cartan matrix for an $A_{N-1}$ Dynkin diagram.
When we resolve the $\BZ_N$ singularity along the whole fixed 
curve $C$, we obtain locally a product of the curve $C$ and 
the chain of spheres we have just described. The resolution 
of a single fixed curve $C$ shifts the Hodge 
number $h^{21}$ by $(N-1)g$ with $g$ being the genus of the curve 
$C$ \cite{Bat}. This can be understood by sending a one-cycle of
the curve $C$ to the three-cycle swept out by the
spheres fibered over the one-cycle \cite{CleHar}. (See also \cite{KaMoPl}
for a discussion of singular curves in the context of
gauge symmetry enhancement in type II theories.)
\bigskip 

Let us turn back to the Gepner models and compare what we have
learned about the resolution of singularities with the appearance
of $(c,c)$-fields in the twisted sectors. As we have pointed out 
before, the Landau-Ginzburg description of minimal models involves 
an identification of the coordinate $z_i$ with the chiral primary 
field $\Phi^{i;l}_{m,s} = \Phi^{i;1}_{1,0}$ in the $i^{th}$ minimal 
model. The subgroup 
$\BZ_K \subset \Gamma$ acts on the latter by multiplication with the 
phase $\exp(2\pi i/(2k_i +4))$. Note that the number $N$ must be 
contained as a factor in $K$ since each of the numbers $\omega_3, 
\omega_4, \omega_5$ is contained in $K$. Hence, we obtain an action of 
$\BZ_N \subset \BZ_K$ by 
$$ \Phi^{i;1}_{1,0} \ \rightarrow \ e^{\frac{2\pi i}{2k_i+4}\frac{K}{N}} 
\ \Phi^{i;1}_{1,0} \ = \ e^{\frac{2\pi i}{N} \omega_i} \  \Phi^{i;1}_{1,0}
  \ = \ \eta^{\omega_i}  \  \Phi^{i;1}_{1,0}\ \ ,  $$
where $\eta = \exp(2\pi i/N)$. This is precisely the transformation law
of the coordinate $z_i$ under the action of the $\BZ_N$ that is 
responsible for the singularity in the weighted projective space.       
Hence, we may identify the subgroup $\BZ_N \subset \BZ_K \subset 
\Gamma$ of the orbifold group with the geometrically acting group 
on our Calabi-Yau manifold $M$. 
\smallskip

Combining the last paragraph with the previous discussion we can 
now formulate the relation between the singularities and $(c,c)$-fields
more precisely \cite{LySch,FKSS}. 
In fact, we expect to find $g$ of the $(c,c)$-fields
with $q=1=\bar q$ in each of the $\nu = 0$ mod $K/N$-twisted sectors. 
As there exist $N-1$ such sectors (the untwisted sector $\nu =0$ is 
not included), the number of new $(c,c)$-fields matches the
shift of the Hodge number that we attributed to the desingularization 
of a single $\BZ_N$-singular curve $C$. 
\smallskip

While these new $(c,c)$-fields possess the same total right- and 
left moving U(1)-charge, i.e.\ $q = \bar q$ by construction, the 
U(1)-charges in the individual minimal models can be different. 
This may be seen from the obvious property $q^{(i)} - \bar q^{(i)} 
= \nu/(k_i+2)$ mod $1$ of fields in the $\nu$-twisted sector
(it holds for $\nu_i =0$). The right hand side of these equations 
is non-zero unless $2 \nu$ can be devided by all the $2k_i +4$. 
But this requires $2 \nu = K$ and hence we conclude that only 
the new $(c,c)$-fields in the  $\nu = K/2$-twisted sector have 
equal left- and right-moving charges $q^{(i)}$ = $\bar q^{(i)}$ 
in all the individual minimal models. It is easy to see that all 
such fields lie on short orbits of our orbifold group $\Gamma$. 
\smallskip

The boundary states we discuss in this work satisfy A-type boundary 
conditions in each minimal model rather than A-type boundary 
conditions for the diagonally embedded superconformal algebra only. 
This means that generalized coherent states in our boundary 
states are necessarily based on primary fields satisfying the 
condition $q^{(i)}$ =$\bar q^{(i)}$. Hence, all we can hope for 
in the following is to find boundary states which are charged 
under the new $(c,c)$-fields in the $\nu =K/2$-twisted sector,
i.e.\ under the $(c,c)$-fields that arise from resolving a 
$\BZ_2$-singularity along a fixed curve $C$.    
\bigskip

\noindent
{\sc Example:} The degree eight Fermat CY space $\BP_{1,1,2,2,2}[8]$ 
corresponds to the choice $(k=6)^2(k=2)^3$ for the minimal models 
used in Gepner's construction. The surface has only one $\BZ_2$ 
singularity over a curve $C$ of genus $g=3$. Resolving this 
singularity shifts the Hodge number $h^{21}$ by 3. It is not 
difficult to list all the $(c,c)$-fields with $q=1 =\bar q$ in 
this case. Except from those in the untwisted sector, there 
appear 3 additional such fields in the $\nu = 8$-twisted sector
with labels for their left-movers being given by  
$$ (3,3,0) \times (3,3,0) \times (1,1,0) \times (0,0,0) \times 
(0,0,0)$$
and by similar expressions obtained through a  permutation of the 
last three factors. Obviously, these fields are all on short orbits 
of the group $\Gamma = \BZ_{16} \times \BZ_2^5$. 

\section{Twisted D-branes in Gepner models}

We have argued in the previous section that a typical Gepner model 
has $(c,c)$-fields in the twisted sectors. Geometrically, this situation 
corresponds to the resolution of singularities inherited from the 
embedding projective space. D-branes can wrap the extra homology 
cycles on the resolved manifold. Therefore, we expect that there 
are boundary states containing Ishibashi states built on the $(c,c)$ 
fields in twisted sectors of the models. This is necessary in order
to have D-branes which are charged under all the possible massless  
RR potentials in the theory. Since locally the singularities we 
shall be concerned with look like $\BZ_2$ orbifolds of flat space, 
we will first review the construction of boundary states for 
fractional branes in flat space. This will help to motivate the 
procedure that allows to find twisted sector boundary states in 
Gepner models. 

\subsection{Twisted boundary states in flat space}

D0 branes at orbifold singularities of flat space have been discussed 
from the open string point of view in \cite{DM}. There are different 
types of branes depending on the representation of the orbifold group 
on the Chan-Paton factors. The group $\BZ_2$ has two inequivalent 
irreducible one-dimensional representations which are distinguished
by a sign. These two representations lead to two types of fractional 
branes. In fact, to each of the two representations one can associate 
a boundary state \cite{ej, HINS, BCR, BER} 
(see also \cite{AffOsh} for a related 
construction in a different context) for branes sitting at the origin
$z = (z_1,z_2) = (0,0)$. They are given by 
$$
|D(z=0);\pm\rangle  \  = \ \frac{1}{\sqrt{2}} \, |D(z=0)\rangle \, \pm \,  
\sqrt{2} \, |D(z=0) \rangle\rangle^{\rm tw}\  .
$$
Here, $|D(z=0)\rangle\rangle^{\rm tw}$ is an Ishibashi state obtained 
from the twisted sector of the $\BZ_2$-orbifold theory. After modular 
transformation, the transition amplitude between two such twisted 
boundary states acquires the form 
\begin{eqnarray*}
Z_{++} \ = \ Z_{--} (q) \ &=& \ Z_{|D0\rangle}(q) + Z^{\rm tw}(q) 
\ = \  \tr \left( \frac{1+g}{2} q^{H_{op}}\right) \ \ \ ,\\ \nn
Z_{+-} \ = \ Z_{-+} (q) \ &=& \ Z_{|D0\rangle}(q) - Z^{\rm tw}(q) 
\ = \  \tr \left( \frac{1-g}{2} q^{H_{op}}\right) \ \ \ ,
\end{eqnarray*}
where $g$ is the non-trivial element of $\BZ_2$ acting on both 
Chan-Paton labels and oscillators. The twisted part $|D(z=0)\rangle 
\rangle^{\rm tw}$ of the boundary state is responsible for the term 
$Z^{\rm tw}(q)$ in the open string partition function in which the 
group element $g$ is inserted. 
\smallskip

If we add the two boundary states $|D;+\rangle$ and $|D;-\rangle$ 
we obtain a boundary state without contributions from twisted 
sectors. This state comes with the regular representation of the 
orbifold group on Chan-Paton labels and it can be obtained directly 
by the projection method applied to the boundary state $|D(x=0)\rangle$
of the theory on the two-fold cover of the orbifold space. Turning the 
argument around, we see that boundary states at fixed points obtained 
with the projection method can be further decomposed by taking
into account twisted sectors. This is the strategy we will now
follow in our discussion of twisted boundary states for Gepner 
models. 

\subsection{Twisted boundary states in the Gepner model}

In this section we will construct additional boundary states for the
Gepner model, which are charged under the twisted sector states
in short orbits of the orbifold group $\Gamma$. The method will 
be similar to the one used in flat space. The role of the state 
$|D(x=0)\rangle \rangle^{\rm tw}$ is played Ishibashi states
built from states in $K/2$-twisted sector of the Gepner models. 
Our aim is to improve the boundary states associated with short 
orbits of $\Gamma$ by adding twisted sector Ishibashi states. 
\medskip

There are a few general statements one can make about this 
procedure whenever $\BZ_2$ appears as the maximal stabilizer 
subgroup of the theory. In this case, the short orbits have
length $K2^{r-1}$. The application of the projection (\ref{prorb})
leads to two equal pieces in the projected sum. Each of the
two pieces can be interpreted as a sum over elements of the
orbifold group modulo the stabilizer. It means that the boundary 
states we started with are not elementary but rather have to be 
replaced by a linear combination of the projected state and 
Ishibashi states from the $\nu = K/2$ twisted sector,
i.e.\ 
\beq
|\Lambda_s,\Xi_s\rangle_\proj \ \mapsto \ \beta \ |\Lambda_s,
\Xi_s\rangle_\proj  \pm \sum_{\lambda_s, \mu_s} 
\alpha_{\lambda_s, \mu_s}\   |\lambda_s, \mu_s\rangle\rangle^{{\rm tw}},
\label{impr} \eeq
where 
and $|\lambda_s, \mu_s \rangle\rangle$ is an Ishibashi state built 
on a short orbit primary labeled by $\lambda_s$ and $\mu_s$.
The coefficients $\alpha$ and $\beta$ in the boundary state have 
to be chosen in such a way that the boundary state leads
to a consistent open string spectrum. We will see that we can easily
get information on $\beta$ from general considerations, whereas
$\alpha$ is model dependent.
\smallskip 

Let us start the discussion by computing the open string spectrum
in the case that one boundary state is given by the improved
short orbit boundary state and the other one is given by a
projected long orbit boundary state. In this case, we can get conditions
on $\beta$ only, since the twist fields are orthogonal to all
fields of the untwisted sector. The open string sector
is determined by the fusion rules of the elements of the short 
orbit  with those of the long orbit (cf.\ eq.\ (\ref{Zproj}) above). 
According to our general discussion following eq.\ (\ref{Zproj}), 
the overall multiplicity of the fields propagating in the open 
string sector is $2\beta$. Therefore, minimal normalization
suggests to pick
$$
\beta \ = \ \frac{1}{2}.
$$
Of course one has to check that this choice gives consistent 
open string spectra when tested against other short orbit 
states of the form (\ref{impr}). We will do this in the 
explicit examples below.
\medskip

To prepare for the discussion of Gepner models, we determine 
the improved boundary states in a single $\CN =2$ minimal model, 
where we mod out by the group generated by the current $(0,1,1)$. 
We pick a minimal model with $k=0$ mod $4$. In this case, 
the length of a generic orbit  of the current $(0,1,1)$ is
$2k+4$ and  there is a short orbit for $l= k/2$. As we outlined 
above, our plan is to add Ishibashi states $|k/2, m+\nu,s+\nu
\rangle\rangle^{\rm tw}$ built on twisted sector fields to the 
short orbit boundary states. Our ansatz for the improved 
boundary states is 
\begin{eqnarray}\label{prefac}
\ket{\frac{k}{2}, M, S} &=& \frac{1}{2}\ \ket{\frac{k}{2}, M, S}_\proj \ + 
\\ \nn &&  + \ \tilde\alpha\ 
\frac{1}{2^{1/4}(k+2)} 
\sum_{\nu=0}^{k+1} \sum_{m,s}^{ev} (-1)^\nu
e^{i\pi m\frac{M+\nu}{k+2}}\,  e^{i\pi s \frac{S+\nu}{2}}\, 
|\frac{k}{2}, m,s \rangle \rangle^{\rm tw}\ \ ,
\end{eqnarray}
where the sum over $m$ and $s$ is constrained to $m+s=even$.
We allow for an overall factor $\tilde\alpha$ in front
of the twisted part of the boundary state.  It will 
be adjusted later so that we get a consistent open string 
spectrum. To this end, we have to compute the transition
amplitude between the twisted parts of the boundary state
and to perform a modular transformation. This results in 
the following expression for the contribution $Z^{\rm tw}$ 
of the twisted sector to the full open string partition 
function : 
\beq
Z^{\rm tw} \ = \ \tilde\alpha^2 \frac{2}{k+2} \sum_{\nu=0}^{k+1} 
\sum_{l',m',s'}^{ev}  (-1)^\nu
\sin \pi \frac{l'+1}{2} \ \delta^{(2k+4)}_{\nu+m'+\frac{k+2}{2}(\nu +s')}
\ \delta^{(2)}_{\nu+s'} \chi_{l',m's'}(q)\ \ . 
\eeq
From our previous discussion on the minimal normalization of $\beta$ 
we know that the untwisted part of the full partition function 
computed from the state (\ref{prefac}) has a factor $1/2$ standing 
in front of each character. We pick $\tilde\alpha$ in such a way that 
there is a factor of $1/2$ in front of the twisted contribution as well. 
This means that we take $\tilde\alpha$ to be
\beq\label{alpha}
\tilde\alpha^2\ = \ \frac{k+2}{4}\ \ . 
\eeq
For this choice to be consistent, all characters in the full 
partition function should come with integer coefficients. The 
twisted sector gives negative contributions for $l'=2 $ mod $4$, 
positive contributions for $l'=0 $ mod $4$ and no contributions 
for $l'$ odd. With our choice of $\tilde\alpha$, all contributing 
characters come with a factor  $\pm 1/2$ in front. 

From the untwisted part of the partition function we 
obtain characters with even $l'$. As we mentioned before, 
their coefficient is $1/2$. Characters with odd $l'$
are forbidden by the fusion rules. Putting everything 
together, we have shown that the sum of the twisted and
the untwisted part of the partition function contains
the characters with $l'=0$ mod $4$ with an integer 
coefficient. The characters with $l'=2 $ mod $4$ get 
subtracted, so that they are not propagating in the 
open string sector.

Equation (\ref{alpha}) for $\tilde \alpha$ admits the choice of a 
sign similar to the flat space case. In our general discussion,
this corresponds to a character of the stabilizer subgroup.
Since our orbifold group is $\BZ_2$, the latter reduces to 
the choice of a sign. The open string partition function has 
an opposite sign in front of the twisted part in the case that 
the two boundary states have opposite signs. In this case, the
characters with $l'=0$ mod $4$ get removed from the partition
function and those with $l'=2$ mod $4$ survive. This can
be summarized in the following formulas
\begin{eqnarray}
Z_{++} \ = \ Z_{--} &=& \sum_{\nu=0}^{k+1} \sum_{l', m', s'}
(-1)^\nu \, \delta^{(4)}_{l'}\,  \delta_{\nu+m' +\frac{k+2}{2} (\nu +s')}
\, \delta^{(2)}_{\nu+s'} \ \chi_{l', m', s'}(q) \nn \\[3mm] 
Z_{+-}\ = \ Z_{-+} &=& \sum_{\nu=0}^{k+1} \sum_{l', m', s'}
(-1)^\nu \, \delta^{(4)}_{l'+2} \, \delta_{\nu+m' +\frac{k+2}{2} (\nu +s')}
\, \delta^{(2)}_{\nu+s'} \ \chi_{l', m', s'}(q) \ \ . \nn
\end{eqnarray}
\bigskip

Let us now discuss short orbit states in the full Gepner model.
The length of a short orbit is $K/2$. In the individual minimal
models, this means that $l_i = k_i/2$ in the case that $2k_i+4$,
the orbit length of the single minimal model, does not divide
$K/2$. In the case that $K/2$ is a multiple of the orbit length
of a minimal model, an arbitrary primary can be chosen in that
particular model. If the number of minimal models, for which we need
$l_i =k_i/2$, is $r'$, the factors in the minimal models can be
ordered in such a way that these minimal models are the factors
$1, \dots, r'$. This means that the labels of the short orbit
boundary state are
\beq
\Lambda_s \ = \ (\frac{k_1}{2}, \dots, \frac{k_{r'}}{2}, L_{r'+1}, 
\dots, L_r)
\eeq
with arbitrary labels $\Xi$. 
Our general discussion of short orbit states and
the discussion of the single minimal model in this section motivates
the following ansatz for a short orbit boundary state in the full
Gepner model
\begin{eqnarray*}
\ket{\Lambda_s, \Xi_s} &=& \frac{1}{2} \ \ket{\Lambda_s, 
\Xi_s}_{\proj}\ +  \\[2mm] && \hspace*{-15pt} + \ \alpha\, c\, N
\sum_{\nu=0}^{\frac{K}{2}-1 }\sum_{\nu_j =0,1}
\sum_{\mu} \sum_{l_{r'+1},\dots,l_r} \ \ (-1)^{\nu} \ (-1)^{\frac{s_0^2}{2}}
\ \ \prod_{j=r'+1}^r
\frac{\sin(l_j, L_j)_{k_j}}{\sqrt{\sin (l_j,0)_{k_j}}} \times \\[3mm]
&& \hspace*{-15pt} \times
\prod_{j=1}^{r} e^{i\pi\frac{M_j+\nu}{k_j+2}m_j} 
e^{-i\pi \frac{s_j}{2}(S_j+\nu+2\nu_j)} 
e^{-i\pi \frac{s_0}{2}(S_j+\nu-2\sum \nu_j)}
\kket{\frac{k_1}{2}, \dots, \frac{k_{r'}}{2}, l_{r'+1}, 
\dots, l_{r}; \mu }^{{\rm tw}}\ ,\\[5mm] 
& & \mbox{\phantom{xxxxxx} where} \ \ \ c\ =\ \lb \sqrt{2}\prod_j 
 \sqrt{\sqrt{2}  (k_j+2)} \rb^{-1}
\end{eqnarray*}
and $N=1/\sqrt{K2^r}$. 
To determine the prefactor $\alpha$ of the twisted part of the
boundary state, we compute the transition amplitude between the boundary
state and itself and modular transform to the open string sector.
This leads to the following result for the twisted part of
the partition function
\begin{eqnarray*}
Z^{\rm tw}&=&
\alpha^2 \, \frac{1}{2}\  \frac{1}{K2^r} 
\sum_{\nu}\sum_{\nu_j} \sum_{\lambda', \mu'} (-1)^\nu 
\, \delta^{(4)}_{\nu - 2\sum\nu_j +s_0' - 2} \\[2mm]
&& \prod_{j=1}^{r'} \frac{2}{k_j+2}\ \sin \pi \frac{l_j'+1}{2} \ \ 
\delta_{\nu+m_j' +\frac{(k_j+2)}{2}(\nu+2\nu_j + s_j')}^{(2k_j+4)}
\ \delta^{(2)}_{\nu+2\nu_j +s_j'}\\[2mm]
&&\prod_{j=r'+1}^{r} N_{L_j L_j}^{l_j'} \ \delta^{(2k_j+4)}_{\nu+m_j'}
\ \delta^{(4)}_{2 \nu_j+\nu +s_j'} \ \ \chi_{\lambda', \mu'}(q)\ \ .
\end{eqnarray*}
The computation for the factors $1, \dots, r'$ is like that for the
short orbits in a single minimal model and the computation for
the other factors works exactly as the computation for the
projected boundary states of long orbits.
\smallskip

Given that we have  characters with a prefactor of $1/2$
in the untwisted part of the partition function, the consistent
choice for $\alpha$ is
\beq
\alpha \ = \ \pm\, \ \sqrt{ K 2^{r-r'-1} 
        \prod_{j=1}^{r'}(k_j +2)} \ \ \ . \eeq
Again, there are two different boundary states $\ket{\lambda_s, \Xi_s; 
\pm}$ depending on the sign chosen for $\alpha$. Adding the twisted and
untwisted part of the partition function, we see that
characters with an odd $l_j'$ do not appear in both the untwisted
and twisted part. Those with an even number of $l_j' = 2$ mod $4$
add up and appear with multiplicity one in the total partition
function. Those with an odd number of $l_j'=2$ mod $4$ appear
with a negative sign in the twisted part of the partition function
and get removed from the spectrum in the total partition function.

In the case of the
Gepner model, there is usually more than one short orbit, and
to check the consistency of the boundary states constructed above
we have to compute the partition functions for all combinations
of boundary states. The short orbits differ in the choice of
$L_j$ in the minimal model factors which are not restricted to
be on short orbits and in the $\Xi_s$. The computation of the
partition function for two different boundary states is
not much different from that above. In the $SU(2)$ fusion
coefficients, one of the $L_j$ is replaced by some other label $\tilde 
L_j$ for all $j= r'+1, \dots, r$. Furthermore, one has to 
substitute the sum $m_j' + \nu$ with $m_j' + \nu + M_j -
\tilde M_j$ and similarly $s_j'+\nu$ with $s_j' +\nu +S_j 
- \tilde S_j$.

\section{Conclusions and Outlook} 
\def\cG{{\cal G}}

In this paper, we have constructed A-type boundary states in the
Gepner model, which cannot be obtained by the projection method. 
These boundary states contain Ishibashi states from twisted
sectors. In particular, they carry RR-charge under the RR-fields
in the twisted sector of the Gepner model. Geometrically, this means
that the branes wrap exceptional cycles, i.e.\ cycles coming from 
the resolution of a singularity. It has been shown in examples 
\cite{Scheid} that the rank of the intersection matrix computed 
from projected boundary states equals the number of homology 
three-cycles that do not come from the resolution. This clearly 
demonstrates that the projection method is missing interesting 
boundary states whenever the corresponding CY-manifold develops 
singular curves. Some of these extra boundary states are now 
provided by our construction.

In this paper, we considered only A-type boundary states. B-type boundary 
states in Gepner models have been obtained  in \cite{RSgepner}. It is 
possible to reinterpret the formulas given there in terms of simple 
currents orbifolds. 
\smallskip

For B-type boundary states, the RR-charges associated with $(a,c)$-%
fields become important and replace the $(c,c)$-fields in our discussion
above. It is well known that mirror symmetry relates these two different 
types of bulk fields, i.e.\ the space of $(a,c)$-fields on the original 
3-fold $M$ gets mapped into the $(c,c)$-fields on the mirror $W$ and 
vice versa. Let us recall that mirror symmetry of closed string theories
can be understood as an orbifold construction \cite{GrePle} with the 
group 
$$
  \cG \ =\ \left(\prod_{i=1}^r  \cG_i \right) / \Gamma'   
\ \cong \ \left(\prod_{i=1}^r \ \BZ_{k_i + 2}\right) / 
  \BZ_{\frac{K}{2}} 
$$  
where $\cG_i$ is the group generated by the element $(2; 0,\dots,
m_i=2,\dots,0;0,\dots,s_i=2,\dots,0)$ and $\Gamma'$ is the intersection 
between the orbifold group $\Gamma$ and the product of groups $\cG_i$ 
in the nominator. Most importantly, this orbifold construction allows
to compute the bulk partition function for the theory on the 
mirror $W$. 
\smallskip

Following our general discussion above, one can extend 
these constructions of the mirror theory to obtain A-type boundary 
states on $W$ through an orbifold procedure with the group $\Gamma 
\times \cG$. Obviously, the resulting spectra in the open string 
sector will be organized in terms of orbits of this orbifold group. 
On the other hand, orbits of $\Gamma \times \cG$ do show up in the 
explicit formulas for open string spectra of B-type boundary states 
in the string theory on the original 3-fold $M$ \cite{RSgepner}. 
The fact that the partition functions of $\Gamma \times \cG$-%
projected (A-type) D-branes on $W$ coincide with those found in 
\cite{RSgepner} for B-type branes on $M$ confirms that the orbifold 
description of mirror symmetry extends to open strings (cp.\ also 
\cite{OoOzYi,GutSat1}). 
\smallskip

In the detailed analysis of the $\Gamma \times \cG$-orbifold theories 
one finds a rich pattern of orbits with short orbits of various 
lengths appearing for a typical model. This signals the existence 
of additional boundary states which are not obtained by projection 
from the tensor product theory. In principal, these states can be 
constructed along the lines of Section 5 above. Only technically 
this becomes more involved due to the possibly complicated orbit 
structure. The resulting B-type boundary states on $M$ are charged 
under fields which arise from desingularization of singular points 
and of singular curves.  
\smallskip

In \cite{DiaRom, Scheid, KLLW} B-type boundary states have been
compared with vector bundles on Calabi-Yau manifolds, which are
elliptic \cite{DiaRom} or K3 \cite{Scheid, KLLW} fibrations.
For short orbit states it was found that there are a number of
vacua propagating. The interpretation of \cite{DiaRom} was
that these boundary states do not correspond to elementary
branes. This is indeed supported by our analysis. The elementary
short orbit states should contain a twisted part.
\medskip

Let us remark that our techniques can also be applied to 
boundary states in bulk theories 
with a D-type modular invariant. Boundary states for such theories 
have been discussed recently in \cite{NakNoz}. In the context of
simple current constructions, D-even modular invariants can be
constructed by modding out an additional $\BZ_2$ current in a 
minimal model. The open string sector is then organized
in orbits of this current.
\bigskip

The CFT-construction we have used in this work allow to obtain 
a large number of boundary states extending the set of states 
which were known before, if the underlying Calabi-Yau 3-fold
possesses $\BZ_2$-singularities along curves. It is interesting 
to point out once more that the geometric scenario of Section 4 
suggests the existence of many other important D-brane states
that are charged under $(c,c)$-fields from the $K/N$-twisted 
sectors when $N\neq 2$. The latter come with the desingularization 
of a $\BZ_{N\neq 2}$-singularity over a curve $C$ on $M$. 
While boundary states charged under such $K/N$-twisted $(c,c)$-%
fields satisfy A-type gluing conditions for the diagonally 
embedded superconformal algebra, it would be inconsistent 
with the U(1) charges of the corresponding $(c,c)$-fields to 
require such A-type gluing conditions for each minimal model
separately. But so far, all treatments of D-branes in Gepner 
models assumed the strong version of the gluing condition 
which preserves all the individual superconformal algebras. 
It remains an interesting open problem to relax this 
requirement and, in particular,  to construct $D$-branes 
associated with the resolution of $\BZ_{N\neq 2}$ singularities
over curves $C$ in $M$.       

\medskip
\newpage

\noindent{\bf Acknowledgements:} We would like to thank
M.\ Bianchi, R.\ Blumenhagen, M.\ Douglas, H.\ Liu, 
A.\ Recknagel, C.\ R\"omelsberger, R.\ Schimmrigk, 
Y.\ Stanev and  S.\ Theisen for very useful discussions.
V.S.\ is grateful to the Rutgers string group for their 
hospitality. I.B.\ thanks the II. Institut f\"ur 
Theoretische Physik for supporting her stay at Hamburg.
We also want to thank A.\ Recknagel for his helpful 
remarks on the manuscript.   

\newpage
\bibliography{gepcorl}
\bibliographystyle{utphys}

\end{document}